\documentclass[prd,aastex,amsmath,amssymb,twocolumn,widetext,floatfix,aps,showpacs,longbibliography]{revtex4-2} 

\usepackage{graphicx}
\usepackage{bm}
\usepackage{textcomp}
\usepackage{xcolor}
\definecolor{myblue}{rgb}{0.0, 0.0, 0.6}
\usepackage{hyperref}
\hypersetup{
  colorlinks = true,
  citecolor  = myblue,
  linkcolor  = myblue,
  urlcolor   = myblue
}

\begin{document}
\title{
  Feasibility study for precisely measuring the EIC ${}^3$He beam polarization with the Polarized Atomic Hydrogen Gas Jet Target polarimeter at RHIC
}%
\author{A.~A.~Poblaguev}\email{poblaguev@bnl.gov}
\affiliation{%
 Brookhaven National Laboratory, Upton, New York 11973, USA
}%
\date{December 12, 2022}
\begin{abstract}
  \noindent The Polarized Atomic Hydrogen Gas Jet Target polarimeter (HJET) is used to measure the absolute proton beam polarization, $\sigma_P^\text{syst}/P\!\lesssim\!0.5\%$,  at the Relativistic Heavy Ion Collider. Here I consider the possibility of employing HJET to measure the ${}^3\text{He}$ ($h$) beam polarization at the Electron-Ion Collider (EIC). The dominant contribution to the ratio of the $h^\uparrow{p}$ and $p^\uparrow{h}$ analyzing powers, which is needed for such measurements, can be easily calculated using well-known values of the proton and helion magnetic moments, but some corrections should be applied to achieve the required accuracy. It was found that corrections due to absorption and ${}^3\text{He}$ breakup effectively cancel in the ratio and a correction due to hadronic spin-flip amplitudes can be derived from the proton beam measurements. As a result, the anticipated systematic uncertainty in the measured ${}^3\text{He}$ beam polarization can satisfy the EIC requirement $\sigma_P^\text{syst}/P\!\lesssim\!1\%$.
\end{abstract}
 
\maketitle

\section{Introduction}

High energy, 41--275\,GeV, polarized proton and helion [${}^3\text{He}$ ($A_h\!=\!3$, $Z_h\!=\!2$)] beams are proposed for the future Electron-Ion Collider (EIC) \cite{Accardi:2012qut}. The EIC physics program is being designed assuming precise knowledge of the beam polarization\,\cite{AbdulKhalek:2021gbh}
\begin{equation}
  \sigma_P^\text{syst}/P \lesssim 1\%.
  \label{eq:systEIC}
\end{equation}

Since hadron beams at EIC will be similar to those at the Relativistic Heavy Ion Collider (RHIC),  experience with absolute proton beam polarization measurements at RHIC is very important for the development of hadronic polarimetry at EIC. 

At RHIC, the absolute vertical proton beam polarization is determined by the Atomic Polarized Hydrogen Gas Jet Target (HJET)\,\cite{Zelenski:2005mz}. A critically important feature of the HJET polarimeter is a relatively low-density gas jet target with no walls or windows which allows one to make continuous during the RHIC store measurements in the Coulomb-nuclear interference (CNI) region.

At HJET, the polarized proton beam is scattered off the hydrogen target and recoil protons are counted in the left/right symmetric recoil spectrometer detectors near $90^\circ$ to the beam direction\,\cite{Poblaguev:2020qbw}. The measured beam spin-correlated asymmetry, $a_\text{N}$,  is proportional to the vertical polarization $P$ of the beam
\begin{equation}
  a_\text{N}(T_R)=A_\text{N}(t) P,
\end{equation}
where $t\!=\!-2m_pT_R$ is momentum transfer squared, $T_R$ is the recoil proton kinetic energy, and $m_p$ is a proton mass. The analyzing power $A_\text{N}(t)$ is dominantly defined by the interference of the electromagnetic spin-flip and hadronic nonflip amplitudes. In this approximation and for the CNI scattering, a simple expression for $A_\text{N}(t)$ was obtained in Ref.\,\cite{Kopeliovich:1974ee} (see Fig.\,\ref{fig:AN}). In the precision measurement of $A_\text{N}(t)$ carried out at HJET\,\cite{Poblaguev:2020qbw}, it was found that analyzing power for the forward elastic $p^\uparrow{p}$ scattering was predicted in Ref.\,\cite{Kopeliovich:1974ee} with several percent accuracy.

Due to the spin-flipping polarized target, HJET can be self-calibrated. For the scattering of the identical particles, since the beam and jet spin correlated asymmetries are concurrently measured using the same recoil protons, the beam polarization can be related via
\begin{equation}
  P_\text{beam} = P_\text{jet}\times a_\text{N}^\text{beam}(T_R)/a_\text{N}^\text{jet}(T_R)
  \label{eq:PolBeam}
\end{equation}
to the jet one $P_\text{jet}\!\approx\!0.96\!\pm\!0.001$, which is monitored by the Breit-Rabi polarimeter\,\cite{Zelenski:2005mz}.

In previous RHIC polarized proton Runs 15 ($E_\text{beam}\!=\!100\,\text{GeV}$) and 17 (255\,GeV), the beam polarization of about $P_\text{beam}\!\sim\!55\%$ was measured with a systematic error of $\sigma^\text{syst}_P/P\!\lesssim\!0.5\%$ and a typical statistical uncertainty of $\sigma_P^\text{stat}\!\approx\!2\%$ per 8\,h RHIC store\,\cite{Poblaguev:2020qbw}. Also, the elastic ${pp}$ analyzing power was precisely determined at both beam energies\,\cite{Poblaguev:2019saw}.

For EIC ${}^3\text{He}$ beam polarimetry, an obvious suggestion is to employ a polarized ${}^3\text{He}$ gas target (He3J) and, thus, utilize the advantage of the HJET method (\ref{eq:PolBeam}), which does not require detailed knowledge of the analyzing power. However, to implement this idea, new experimental techniques should be developed for He3J to allow measurements in the CNI region as well as for precision monitoring, $\sigma_P/P\!\ll\!1\%$, of the helion target polarization. Also, backgrounds in He3J may significantly differ from those of HJET. In other words, it is not proved yet that the low systematic uncertainties achieved at HJET would apply in the He3J case.

In this paper, the feasibility of precisely measuring the EIC ${}^3\text{He}$ beam polarization using HJET is investigated.

Since 2015, HJET routinely operates in the RHIC ion beams. The recoil spectrometer performance was found to be very stable for a wide range of ion species ($d$, O, Al, Zr, Ru, Au) and over the beam energies used (3.8--100 GeV/nucleon)\,\cite{Poblaguev:2020qbw}.

Generally, $A_\text{N}^\text{beam}\ne A_\text{N}^\text{jet}$ for nonidentical beam and target (jet) particles. Therefore, for $hp$ scattering, Eq.\,(\ref{eq:PolBeam}) should be adjusted by the analyzing power ratio:
\begin{align}
  &P_\text{beam} = P_\text{jet}\,a_\text{N}^\text{beam}/a_\text{N}^\text{jet}\times{\cal R},
  \label{eq:PolBeamR} \\
  &{\cal R}=\frac{\mu_p-1}{\mu_h/Z_h-m_p/m_h}\times\left[1+\text{corr}\right].
\end{align}
In the leading order approximation, the ratio can be expressed\,\cite{Buttimore:2009zz} via magnetic moments of a proton, $\mu_p$, and a helion, $\mu_h$. However, anticipated corrections, for example, due to the hadronic spin-flip amplitude and due to a possible breakup of the beam ${}^3\text{He}$ should be considered.

In this paper, numerical analysis of the corrections was done for 100\,GeV/nucleon ${}^3\text{He}$ beam scattering off proton target in CNI region, $-t\!<\!0.02\,\text{GeV}^2$. It was found that the corrections are small and well controllable, which suggests that the EIC requirement (\ref{eq:systEIC}) can be satisfied in the HJET measurements.

\section{Helicity amplitudes for forward elastic scattering}

Polarization phenomena in the CNI elastic scattering of the transversely polarized spin-1/2 beam particle off the spin-1/2 target (generally non-identical) is greatly predetermined by two helicity amplitudes\,\cite{Buttimore:1998rj,PhysRevD.18.694,*PhysRevD.35.407,Kopeliovich:1974ee}, nonflip 
\begin{equation}
  \phi_+(s,t) = \frac{\langle+\!+|{\cal M}|+\!+\rangle\:+\:\langle+\!-|{\cal M}|+\!-\rangle}{2} 
  \label{eq:phi+} 
\end{equation}
and spin-flip
\begin{equation}
  \phi_5(s,t) = \langle+\!+|{\cal M}|+\!-\rangle. 
\label{eq:phi5}
\end{equation}
The subscripts used follow the notations of Ref.\,\cite{Buttimore:1998rj}. Both amplitudes can be decomposed to the hadronic and electromagnetic parts
\begin{equation}
  \phi(s,t) \to \phi^\text{had}(s,t) + \phi^\text{em}(s,t)\times e^{i\delta_C(s,t)},
\end{equation}
where $\delta_C$ is the Coulomb phase\,\cite{Cahn:1982nr,Kopeliovich:2000ez}, and $s$ and $t$ are center of mass energy squared and momentum transfer squared, respectively. 

The optical theorem fixes the imaginary part of the forward, $t\!\to\!0$, hadronic nonflip amplitude:

\begin{equation}
  \phi^\text{had}_+(s,t) = \left[i+\rho^{bt}(s)\right]\times s\sigma^{bt}_\text{tot}(s)e^{B^{bt}(s)t/2}/8\pi,
  \label{eq:phi_had}
\end{equation}
where $\sigma_\text{tot}$ is the total cross section, $B$ is the \lq\lq{slope,}\rq\rq and indexes $b$ and $t$ specify beam and target particles, respectively. The hadronic spin-flip amplitude is commonly referred to by a dimensionless complex parameter $r_5\!=\!R_5\!+\!iI_5$\,\cite{Buttimore:1998rj} defined as
\begin{equation}
  \phi_5^\text{had}(s,t)/\text{Im}\,\phi_+^\text{had}(s,t) =  
  (\sqrt{-t}/m_p)\times r_5^{bt}(s).
\end{equation}
Without losing generality, it will be assumed that only the beam particle is polarized.
For immediate estimates, $r_5$  can be approached\,\cite{Poblaguev:2019saw} by
\begin{equation}
  |r_5|\approx0.02.
  \label{eq:r5}
\end{equation}

The electromagnetic amplitudes were calculated in Ref.\,\cite{PhysRevD.18.694,*PhysRevD.35.407}. Neglecting Dirac and Pauli form factors, they can be written\,\cite{Buttimore:2009zz,Poblaguev:2019vho} as
\begin{align}
  \phi_+^\text{em}(s,t)/\text{Im}\,\phi_+^\text{had}(s,t) &=%
  t_c/t+{\cal B}^\text{nf},  \\
  \phi_5^\text{em}(s,t)/\text{Im}\,\phi_+^\text{had}(s,t) &=%
  \left(t_c/t+\cal{B}^\text{sf}\right)\,\varkappa_b'\sqrt{-t}/m_p \label{eq:em5}
\end{align}
\begin{align}
  t_c &= -8\pi\alpha Z_b Z_t/\sigma_\text{tot}^{bt}, \\
  \varkappa'_b &= \varkappa_b-2m_b^2/s
  \approx \varkappa_b-m_b/m_t\times m_p/E_\text{beam}, \label{eq:kappa'}\\
  \varkappa_b&=\mu_b/Z_b-m_p/m_b.
\end{align}
Here, $\mu_b$ is a magnetic moment in nuclear magnetons of the beam particle ($\mu_p\!=\!2.793$ for a proton and $\mu_h\!=\!-2.128$ for a helion) and $E_\text{beam}$ is the beam energy per nucleon. $\cal{B}^\text{nf}$ reflects the difference between Coulomb $F_\text{em}^\text{nf}(t)$ and hadronic $F_+(t)$ form factors and can be approximated as
\begin{equation}
  {\cal B}^\text{nf}/t_c = \frac{d}{dt}\left[\frac{F_\text{em}^\text{nf}(t)}{F_+(t)}\right]_{t=0}=
  \frac{{\cal B}_a^\text{nf}}{t_c}+\frac{r_b^2+r_t^2}{6}-\frac{B^{bt}}{2},
  \label{eq:Beff}
\end{equation}
where $r_b$ and $r_t$ are rms charge radii of the beam and target particles ($r_p\!=\!0.841(1)\,\text{fm}$\,\cite{Zyla:2020zbs},  $r_h\!=\!1.96(1)\,\text{fm}$\,\cite{Shiner:1995zz,*Morton:2006zz}) and  ${\cal B}_a^\text{nf}$ parametrize the absorptive (i.e., due to the initial and final state inelastic hadronic interactions) correction\,\cite{Kopeliovich:2021rdd}, $F_\text{em}(t)\!\to\!F_\text{em}(t)\exp{\left({\cal B}_at/t_c\right)}$, to the nonflip electromagnetic form factor. To evaluate spin-flip ${\cal B}^\text{sf}$ for the $p^\uparrow{h}$ scattering, one can use, in Eq.\,(\ref{eq:Beff}), ${\cal B}_a^\text{sf}\approx\alpha/2$\,\cite{Poblaguev:2021xkd} and, also, the proton charge radius $r_b\!=\!r_p$ should be replaced by the magnetic one $r_p^\text{M}\!=\!0.851(26)\,\text{fm}$\,\cite{Lee:2015jqa}. Since helion spin is mostly carried by the neutron\,\cite{Friar:1990vx}, the magnetic radius of a neutron $r_n^\text{M}\!=\!0.873(11)\,\text{fm}$\,\cite{Kubon:2001rj} is about the same as for a proton, and a ${}^3\text{He}$ is dominantly in a space symmetric state, one may not distinguish between values of ${\cal B}^\text{sf}$ for $p^\uparrow{h}$ and $h^\uparrow{p}$ scattering. 

\begin{figure}[t]
  \begin{center}
    \includegraphics[width=0.9\columnwidth]{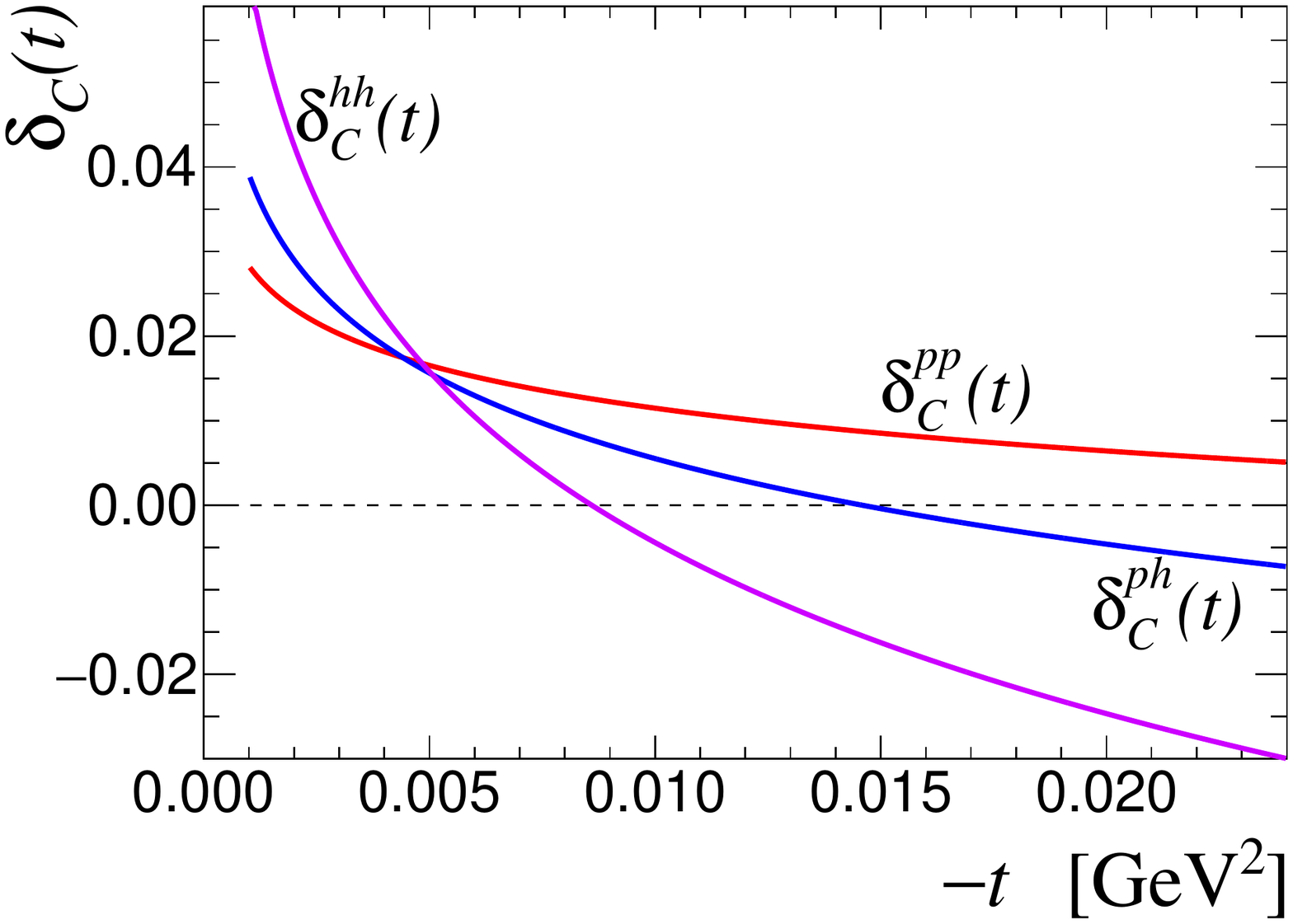}
  \end{center}
  \caption{\label{fig:dC}
    Coulomb phase in elastic proton and/or helion scattering at $E_\text{beam}\!=\!100\,\text{GeV/nucleon}$.
  }
\end{figure}

Coulomb phases for ${pp}$, ${ph}$, and ${hh}$ scattering calculated, in accordance with Ref.\,\cite{Buttimore:1998rj,Poblaguev:2021xkd}, as 
\begin{equation}
  \delta_C(t) = \alpha Z_b Z_t\,\left[%
    \ln{\left|\left(\frac{B^{bt}}{2}\!+\!\frac{r_b^2\!+\!r_t^2}{6}\right)t\right|}\!+\!0.5772\right]
\end{equation}

 are displayed in Fig.\,\ref{fig:dC}.

\section{Analyzing power}

\begin{figure}[t]
  \begin{center}
    \includegraphics[width=0.9\columnwidth]{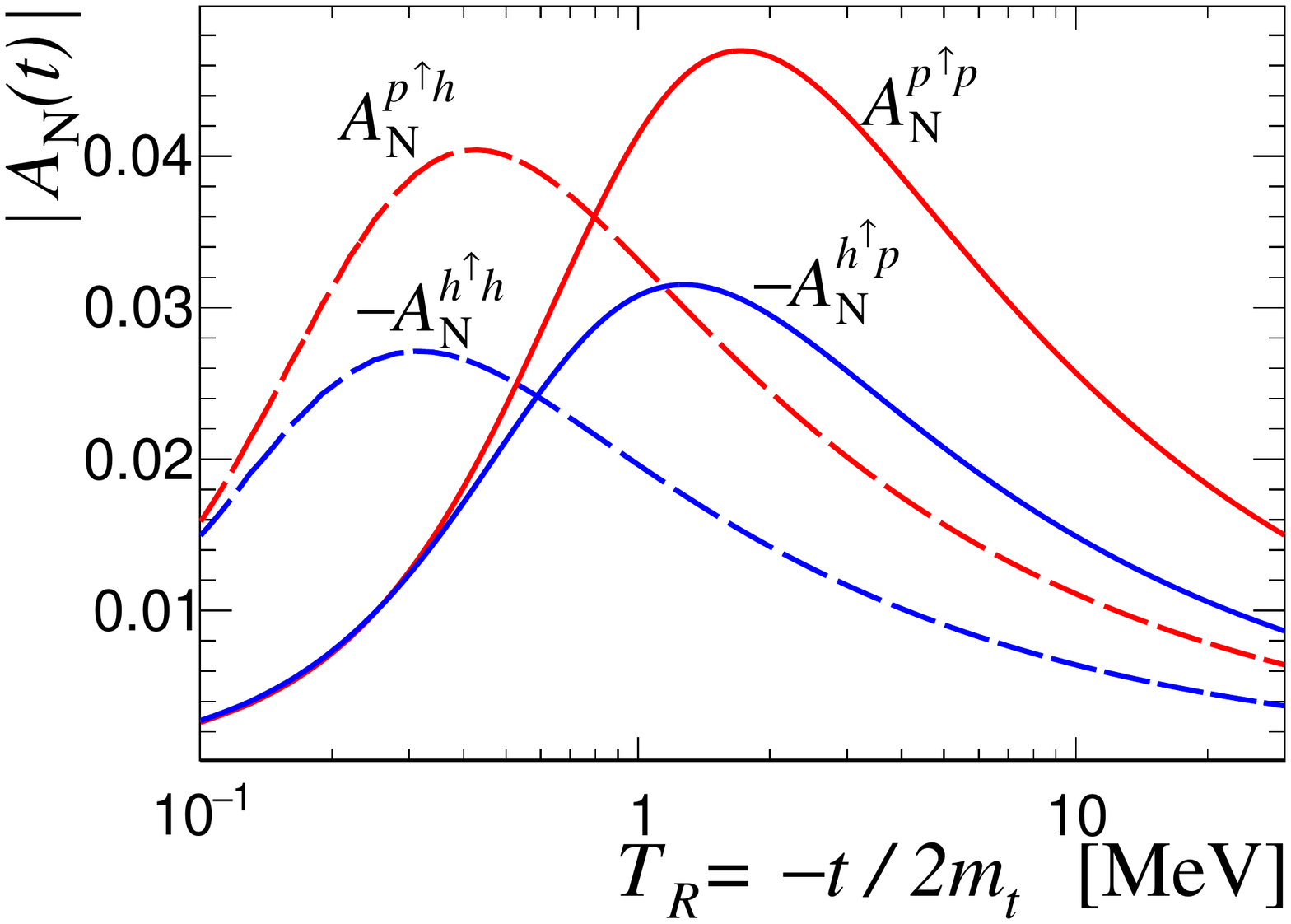}
  \end{center}
  \caption{\label{fig:AN}
    The elastic $p$ and ${}^3\text{He}$ analyzing powers calculated [Eq.\,(\ref{eq:AN_KL})] for $\sigma_\text{tot}^{pp}\!=\!\sigma_\text{tot}^{ph}/0.9A_h\!=\!\sigma_\text{tot}^{hh}/0.8A_h^2\!=\!39\,\text{mb}$. $T_R$ is the recoil (target) particle  kinetic energy.
  }
\end{figure}

 In the considered approximation, the analyzing power is given by
\begin{equation}
  A_\text{N}(s,t) = {-2\text{Im}\left(\phi_5^*\phi_+\right)}/{|\phi_+|^2}.
  \label{eq:AN_ampl}
\end{equation}

The dominant component\,\cite{Kopeliovich:1974ee},
\begin{equation}
  A_\text{N}(t)= \frac{\varkappa_b\sqrt{-t_c}}{m_p}
    \times \frac{(t_c/t)^{1/2}}{(t_c/t)^2 + 1},
  \label{eq:AN_KL}
\end{equation}
can be easily calculated if the total cross section $\sigma_\text{tot}^{bt}$ is known. In this approach, $|A_\text{N}(t)|$ reaches maximum at $t\!=\!\sqrt{3}t_c$. For the proton and/or helion beam and target, the $A_\text{N}(t)$ dependence on the experimentally measured recoil energy $T_R\!=\!-t/2m_t$ is depicted in Fig.\,\ref{fig:AN}.

Analyzing power uncertainties in Eq.\,(\ref{eq:AN_KL}) are about 5--10\%\,\cite{Poblaguev:2020qbw}. To meet EIC requirement (\ref{eq:systEIC}), corrections to Eq.\,(\ref{eq:AN_KL}) should be considered:
\begin{align}
  A_\text{N}(t) &= A_\text{N}^{(0)}(t)\times(1+\xi_0+\xi_1t/t_c),
  \label{eq:alpha}
  \\
  A_\text{N}^{(0)}(t) &= \varkappa_b'\sqrt{-t}/m_pF_\text{cs}(t),
  \label{eq:Acs} \\
  F_\text{cs}(t) &=t_c/t + 2Z_bZ_t\alpha\ln{(t_c/t)}
  \nonumber \\
  &+ \beta_0 + (1\!+\!\beta_1)\,t/t_c + \beta_2\left(t/t_c\right)^2,
  \label{eq:beta} 
\end{align}
where, omitting some small terms,
\begin{align}
  \xi_0 &= -\rho\delta_C(t_c)-2I_5/\varkappa'_b,~~~~~%
  \xi_1={\cal B}^\text{sf}-2R_5/\varkappa'_b,
  \\
  \beta_0 &= -2\left[\rho+\delta_C(t_c)-{\cal B^\text{nf}}\right],~~%
  \beta_1=\rho^2,~~~~%
  \beta_2=0.
  \label{eq:beta1-3}
\end{align}
Here, $A_\text{N}^{(0)}(t)$ is the analyzing power neglecting some small corrections, in particular, due to the hadronic spin-flip amplitude $r_5$. In Eq.\,(\ref{eq:beta}), the term $\beta_2$ is reserved for a possible correction due to the beam ${}^3\text{He}$ breakup.

  It should be noted \cite{Poblaguev:2019vho} that $\rho^\text{meas}\!=\!\rho-{\cal B}_a^\text{nf}$ is actually measured in an experimental study of $d\sigma/dt$ (if absorptive corrections are not considered in the data analysis). Thus, if such a determined $\rho^\text{meas}$ is used to calculate $\beta_0$ then one should set ${\cal B}_a^\text{nf}\!=\!0$ in Eq.\,(\ref{eq:Beff}).
  
Similarly, the spin-flip absorptive correction leads to an effective replacement
\begin{equation}
  r_5 \to r_{5}^\text{meas} = r_5 -\varkappa_b{\cal B}_a^\text{sf}/2,\qquad%
  {\cal B}_a^\text{sf}\approx \alpha/2.
  \label{eq:r5Meas}
\end{equation}

It is assumed that $A_\text{N}^{(0)}(s,t)$ can be unambiguously derived from standalone measurements of the unpolarized elastic cross section $d\sigma/dt\propto|\phi_+(s,t)|^2$. The hadronic spin-flip amplitude contributes only to $\xi_0$ and $\xi_1$.

If the beam polarization $P_\text{beam}$ is known, one can determine hadronic spin flip amplitude parameter $r_5$ from the measured spin asymmetry $a(T_R)$ dependence on $T_R$:
\begin{equation}
  \xi_0 + \xi_1\,T_R/T_c = a(T_R)/P_\text{beam}A_\text{N}^{(0)}(-2m_pT_R)-1.
  \label{eq:r5fit}
\end{equation}
Alternatively, if $r_5$ is predetermined, the beam polarization can be found. For that, it may be helpful to measure beam polarization as function of $T_R$,
\begin{eqnarray}
  P_\text{meas}(T_R) &=& a(T_R)/A_\text{N}(T_R) \nonumber \\
  &\approx& P_\text{beam}\times\left(1-\delta\xi_0-\delta\xi_1\,T_R/T_c\right)
  \label{eq:Pmeas}
\end{eqnarray}
where $\delta\xi_0$ and $\delta\xi_1$ denote systematic errors in the values of $\xi_0$ and $\xi_1$, respectively. Since $\delta\xi_1$ can be determined directly in the linear fit of Eq.\,(\ref{eq:Pmeas}), the systematic uncertainty in the measured beam polarization is actually predefined by $\delta\xi_0$ only:
\begin{equation}
  \delta^\text{syst} P_\text{beam}/P_\text{beam} = -\delta\xi_0.
  \label{eq:dP}
\end{equation}

\section{Systematic errors due to uncertainties in $A_\text{N}^{(0)}(t)$}

Uncertainties in the predetermined $A_\text{N}^{(0)}(t)$ and, consequently, in data fits (\ref{eq:r5fit}) or (\ref{eq:Pmeas}) may depend on a possible error in the value of total cross section $\delta_\sigma\!=\!\delta\sigma_\text{tot}/\sigma_\text{tot}\!=\!-\delta t_c/t_c$ as well as on errors $\delta\beta_i$  in the $d\sigma/dt$ parametrization.  In RHIC Run\,15 (100\,GeV), the resulting effective alterations of the spin-flip corrections $\Delta\xi_0$ and $\Delta\xi_1$ were estimated as
\begin{eqnarray}
  \!\!\!\Delta\xi_0 &=&  0.22\,\delta_\sigma
  +0.49\,\delta\beta_0+0.58\,\delta\beta_1-0.82\,\delta\beta_2 \label{eq:a0},\\
  \!\!\!\Delta\xi_1 &=& 0.09\,\delta_\sigma
  -0.05\,\delta\beta_0+0.08\,\delta\beta_1-0.20\,\delta\beta_2 \label{eq:a1}.
\end{eqnarray}
It should be understood that linearized parametrizations given in Eqs.\,(\ref{eq:a0}) and (\ref{eq:a1}) may strongly depend on the event selection cuts used.

The uncertainties in $A_\text{N}^{(0)}(t)$ may also lead to a non-linearity of Eq.\,(\ref{eq:r5fit}). Considering the effective variation of $\beta_0$ as a measure of the non-linearity, one finds (for Run\,15 data)
\begin{equation}
    \Delta\beta_0 = 4.20\,\delta_\sigma
  -1.00\,\delta\beta_0+2.10\,\delta\beta_1-2.92\,\delta\beta_2 \label{eq:b0},
\end{equation}
where $\Delta\beta_0$ is a correction to the value of $\beta_0$ found in the fit with $\beta_0$ considered as a free parameter. In the HJET data analysis\,\cite{Poblaguev:2020qbw}, this effect was used for an independent experimental evaluation of $\rho$. In terms of measured correction, $\Delta\rho\!=-\!\Delta\beta_0/2$, to the value of $\rho$ assumed, the results can be presented as
\begin{equation}
  \Delta\rho = 0.029(25)|_\text{100\,GeV},~%
  \Delta\rho =-0.019(18)|_\text{255\,GeV}.
  \label{eq:Dbeta0}
\end{equation}
Thus, the value of $\beta_0$ can be used as a free parameter in the spin-correlated asymmetry fit. However, the subsequent contribution of about 2\% to the polarization uncertainty\,(\ref{eq:dP}), in this case, does not allow one to meet the EIC requirement (\ref{eq:systEIC}).  

To parametrize $A_\text{N}^{(0)}(t)$, one can fit a measured differential cross section at low $t$,
\begin{equation}
  d\sigma/dt = \sigma_\text{tot}^2(1+\rho^2)e^{Bt}/16\pi 
  \times F_\text{cs}(t)\,t_c/t.
  \label{eq:dsdt}
\end{equation}
In an experimental study of $d\sigma/dt$, the luminosity calibration is critically important for the precision determination of $\sigma_\text{tot}$, while parameters $\beta_i$ can be extracted from the $t$-dependence fit of Eq.\,(\ref{eq:dsdt}). For elastic ${pp}$ scattering, the fit is significantly simplified because $\beta_2\!=\!0$ and the value of $\beta_1\!=\!\rho^2$ is small.

For $E_\text{beam}\!=\!100\,\text{GeV}$ elastic proton-proton scattering, $B^{pp}\!=\!11.2(2)\,\text{GeV}^{-2}$\,\cite{Bartenev:1973jz,*Bartenev:1973kk}, $\sigma_\text{tot}^{pp}\!=\!38.39(2)\,\text{mb}$ and $\rho^{pp}\!=\!-0.079(1)$. The values of $\sigma_\text{tot}^{pp}$ and $\rho^{pp}$  are taken from a global fit\,\cite{Fagundes:2017iwb} of the ${pp}$ and ${p\bar{p}}$ data. Consequently, $t_c^{pp}\!=\!-1.86\!\times\!10^{-3}\,\text{GeV}^2$ which corresponds to the proton recoil energy of $T_c^{pp}\!\approx\!1\,\text{MeV}$. 

In the absolute majority of experimental evaluations of the proton-proton forward real to imaginary ratio $\rho^{pp}$ at RHIC/EIC energies, the data were analyzed using the electromagnetic form factor in the dipole approximation, $(1-t/\Lambda^2)^{-2}$ with $\Lambda^2=0.71\,\text{GeV}^2$\,\cite{Chan:1966zza}, and disregarding absorptive correction, ${\cal{B}}_a^\text{nf}\!=\!0$. Since actually $\rho^{pp}_\text{exp}\!=\!\rho^{pp}\!-\!{\cal B}_a^{pp}\!-\!(r_p^2/3\!-\!4/\Lambda^2)t_c$ was measured in these experiments, the expression for $\beta_0$ in (\ref{eq:beta1-3}) should be replaced by
\begin{equation}
  \!\!\beta_0^{pp}= -2\left[\rho^{pp}_\text{exp}+\delta^{pp}_C(t_c)%
  + \left(r_p^2/3 - 4/\Lambda^2\right)t_c\right]\!.
\end{equation}
After that, for the ${pp}$ scattering, $d\sigma/dt$ related uncertainties in $A_\text{N}^{(0)}(t)$ and, consequently, in values of $\xi_0$ and $\xi_1$ can be neglected. Nonetheless, $\text{Re}\,r_5$, determined in a hadronic spin-flip amplitude fit\,(\ref{eq:r5fit}), is a little ($\approx\!0.003$) altered by the spin-flip absorptive correction [see Eq.\,(\ref{eq:r5Meas})].

For the helion-proton scattering at the EIC energies, a precision experimental study of the forward $d\sigma^{hp}/dt$ was not done yet.

For a light nucleus $A$, the total $pA$ cross section can be related to the ${pp}$ one as 
\begin{equation}
  \sigma_\text{tot}^{p\text{A}} =
  A\sigma_\text{tot}^{pp}\times\left(1-\delta_\text{A}\right).
\end{equation}
To evaluate the screening parameter $\delta_\text{A}$\,\cite{Glauber:1955qq}, one can approximate the $pA$ forward scattering amplitude $f_\text{A}$ by
\begin{equation}
  f_\text{A} = \sum_{i}{f_i} +
  i{\langle r^{-2}\rangle}_\text{A}\sum_{i\ne j}{f_if_j/2}+\dots, 
  \label{eq:fA}
\end{equation}
where indexes $i$ and $j$ denote the constituent neutrons and protons, and it is assumed that forward ${pp}$ and ${pn}$ amplitudes are equal, $f_n\!=\!f_p\!=\!(i+\rho^{pp})\sigma_\text{tot}^{pp}/4\pi$. Following Ref.\,\cite{Jenkins:1978fv}, one can estimate $\delta_d\!=\!0.051$, $\delta_h\!=\!0.119$, and $\delta_\alpha\!=\!0.246$ for deuteron, helion, and alpha, respectively. 

The result for the deuteron is in good agreement with the value of $\delta_d^\text{exp}\!=\!0.047(5)$ derived from the $\sigma_\text{tot}$ measurement\,\cite{Carroll:1974yv,*Carroll:1975xf} at 100\,GeV, but there is a noticeable discrepancy for ${}^4\text{He}$, $\delta_\alpha^\text{exp}\!=\!0.166(6)$\,\cite{Burq:1980nb}. Therefore, it was assumed that
\begin{equation}
  \delta_h = 0.1\times\left[1\pm(20\text{--}30)\%\right]
  \label{eq:delta_h}
\end{equation}
to depict $A_\text{N}(t)$ in Fig.\,\ref{fig:AN}. 

EIC requirement (\ref{eq:systEIC}) implies the following constraints (considered separately) on accuracy of the $d\sigma^{hp}/dt$ parametrization:
\begin{equation}
  |\delta_\sigma^{hp}|    \!<\! 5\%,\,\:
  |\delta{\beta_0^{hp}}| \!<\! 2\%,\,\:
  |\delta{\beta_1^{hp}}| \!<\! 2\%,\,\:
  |\delta{\beta_2^{hp}}| \!<\! 1\%.
\end{equation}
It may be noted that uncertainty in value of the ${hp}$ cross-section followed from Eq.\,(\ref{eq:delta_h}) is consistent with the constraint on $\delta_\sigma^{hp}$.

In ${pp}$ forward elastic Re/Im ratio experiments\,\cite{Zyla:2020zbs}, the typical statistical accuracy of a measurement was $|\delta\rho|\gtrsim0.01$  or  $|\delta{\beta_0^{pp}}|\!\gtrsim\!2\%$. (However, combined analysis\,\cite{COMPETE:2002jcr} of the all measurements significantly improves accuracy of the determination of  $\beta_0^{pp}$.)  For ${hp}$ scattering, the experimental evaluation of $\beta_0$ is expected to be less accurate, $|\delta{\beta_0^{hp}}|\!>\!|\delta{\beta_0^{pp}}|$ due to the breakup related uncertainties in $\beta_1$ and $\beta_2$.

A theoretical extrapolation of the ${pp}$ values of $\sigma_\text{tot}$, $\rho$, and $B$ to those of ${hp}$, if this is possible to within the required accuracy, may also be complicated by the necessity to precisely evaluate the nonflip absorptive correction to a $hp$ electromagnetic amplitude ${\cal B}_a^{hp}$ and the $^{3}\text{He}$ breakup contributions to $\beta_0$, $\beta_1$, and $\beta_2$.

A detailed investigation of the possibility to improve $d\sigma^{hp}/dt$ parametrization in a theoretical analysis and/or in a dedicated experimental study is beyond the scope of this paper.

\section{${}^3\text{He}$ beam polarization measurement using polarized hydrogen jet target}

For ${}^3\text{He}^\uparrow$ beam scattering off the polarized hydrogen jet, the anticipated problem with the cross section related uncertainties can be eliminated by concurrent measurement of the beam and jet spin asymmetries:
\begin{equation}
  \frac{a_\text{N}^{h}(T_R)}{a_\text{N}^{p}(T_R)} =
  \frac{P_\text{beam}}{P_\text{jet}} \times
  \frac{\varkappa_h'(1\!+\!\omega_\varkappa)-2I_5^{hp}-2R_5^{hp}\,T_R/T_c}%
       {\varkappa_p'(1\!+\!\omega_\varkappa)-2I_5^{ph}-2R_5^{ph}\,T_R/T_c}
  \label{eq:He3Pol}
\end{equation}
Here, indexes ${ph}$ and ${hp}$ relate to the $p^\uparrow h$ and $h^\uparrow{p}$ hadronic spin-flip amplitudes, respectively, and 
\begin{equation}
  \omega_\varkappa(T_R) = -\rho\delta_C - {\cal B}^\text{sf}T_R/T_c + b(T_R)
  \label{eq:kappaCorr}
\end{equation}
where $b(T_R)$ is the effective correction to $\phi_+^\text{had}$ due to the ${}^3\text{He}$ breakup. Since both $r_5^{ph}$ and $r_5^{hp}$ are expected to be small (\ref{eq:r5}), $\omega_\varkappa$ can be omitted in Eq.\,(\ref{eq:He3Pol}) if $|b(T_R)|\!<5\%$. According to the estimate in Ref.\,\cite{Poblaguev:2022hsi}, this condition for $b^{hp}(T_R)$ is fulfilled. Thus, the helion beam polarization derived from Eq.\,(\ref{eq:He3Pol}) can be approximated as
\begin{equation}
  P_\text{meas}(T_R)=P_\text{beam}\times\left(%
  1+\delta\xi_0+\delta\xi_1\,T_R/T_c
  \right)
\end{equation}
where the method dependent systematic errors,
\begin{eqnarray}
  \delta\xi_0 &=&
  2\delta{I_5^{hp}}/\varkappa_h'-2\delta{I_5^{ph}}/\varkappa_p',
  \\
  \delta\xi_1 &=&
  2\delta{R_5^{hp}}/\varkappa_h'-2\delta{R_5^{ph}}/\varkappa_p',
\end{eqnarray}
are predefined by uncertainties in the values of $r_5^{ph}$ and $r_5^{hp}$.

\subsection{Hadronic spin-flip amplitudes in elastic $h^\uparrow p$ and $p^\uparrow h$ scattering}

Currently, theory cannot predict $r_5^{hp}$ from first principles. However, precision measurements of the proton-proton $r_5$ at HJET\,\cite{Poblaguev:2019saw}
\begin{align}
  \text{100\,GeV:~~}&%
  R_5^{pp} = \left(%
  -16.4\!\pm\!0.8_\textrm{stat}\!\pm\!1.5_\textrm{syst}\right)\!\times\!10^{-3},
  \label{eq:R5_100} \\
  &
  I_5^{pp}\, = \left(\;\;%
   -5.3\!\pm\!2.9_\textrm{stat}\!\pm\!4.7_\textrm{syst}\right)\!\times\!10^{-3},
   \label{eq:I5_100} \\
  \text{255\,GeV:~~}&%
  R_5^{pp} = \left(\;\;%
  -7.9\!\pm\!0.5_\textrm{stat}\!\pm\!0.8_\textrm{syst}\right)\!\times\!10^{-3},
  \label{eq:R5_255} \\
  &
  I_5^{pp}\, = \left(\phantom{-}%
   19.4\!\pm\!2.5_\textrm{stat}\!\pm\!2.5_\textrm{syst}\right)\!\times\!10^{-3},
  \label{eq:I5_255}
\end{align}
allow one to consider the possibility of extrapolating $r_5^{pp}\!\to\!r_5^{hp}$ and $r_5^{pp}\!\to\!r_5^{ph}$.

It was shown in Ref.\,\cite{Kopeliovich:2000kz} that at high energy, under a wide range of assumptions, the ratio of the spin-flip to the nonflip parts of a CNI elastic proton-nucleus amplitude is equal to that of proton-proton scattering at the same beam energy (per nucleon). In particular,
\begin{equation}
  r_5^{ph}/r_5^{pp} = 1\pm(\lesssim5\%).
  \label{eq:r5ph_}
\end{equation}
This statement can be qualitatively understood in the $pA$ amplitude approach given by Eq.\,(\ref{eq:fA}). Since the proton-nucleon ratio of the spin-flip and nonflip amplitudes
\begin{align}
  \eta &= f'_n/f_n=f'_p/f_p \nonumber \\
       &= \left(1+\rho^{pp}\right)r_5^{pp}\,\sqrt{-t}/m_p\approx0.002
\end{align} 
is small for the HJET $t$ range, the expression for the spin-flip amplitude $f'_\text{A}$ can be derived by replacing $f_i\!\to\!\eta{f_i}$ in (only) one amplitude in each term of sum (\ref{eq:fA}).  The obvious result, $f'_\text{A}\!=\!\eta f_\text{A}$, can be interpreted as
\begin{equation}
  r_5^{ph}/r_5^{pp} = \left(i+\rho^{pp}\right)/\left(i+\rho^{ph}\right)\approx
  1-i\rho^{pp}\delta_\text{h}.
\end{equation} 
This estimate can be altered by a possible difference between a spin-flip and nonflip values of $\langle{r^{-2}}\rangle_\text{A}$. However\,\cite{Kopeliovich:2000kz}, the subsequent correction to $r_5$, which was included in uncertainty in Eq.\,(\ref{eq:r5ph_}), is small and nearly $A$ independent.

For an unpolarized proton scattering off a nucleus with only one polarized nucleon, one can readily find $r_5^{Ap}\!=\!r_5^{pp}/A$ if $\langle{r^{-2}}\rangle_\text{A}$ is the same for each pair of the constituent nucleons. This is expected to be a good approximation for $h^\uparrow{p}$ scattering, i.e., $r_5^{hp}=r_5^{pp}/3$\,\cite{Buttimore:2001df}, since a ${}^3\text{He}$ is dominantly in a space symmetric $S$ state, in which the constituent protons are restricted by the Pauli principle to be a spin singlet and, thus, the helion spin is carried by the neutron only. In a more detailed analysis of a helion wave function, it was found that the polarization of the neutron in a fully polarized ${}^3\text{He}$ is 87\%, while protons should have a slight residual polarization of $-2.7\%$\,\cite{Friar:1990vx}. So, $r_5^{hp}/r_5^{pp}\!=\!0.27\!\pm\!0.06$, where the uncertainty reflects the discrepancy between two estimates.

Here, the following relations between helion-proton and proton-proton hadronic spin-flip amplitudes will be assumed
\begin{align}
  r_5^{ph} &= \phantom{0.27}\,r_5^{pp} \pm 0.001 \pm i\,0.001, \label{eq:r5ph}\\
  r_5^{hp} &=           0.27\,r_5^{pp} \pm 0.001 \pm i\,0.001. \label{eq:r5hp}
\end{align}

\subsection{Evaluation of the method dependent uncertainties in the ${}^3\text{He}$ beam polarization measurement}

The nonidentical particle factor in the asymmetries ratio (\ref{eq:He3Pol}) is mainly defined by the well-known magnetic moment based ratio $\varkappa_h/\varkappa_p\!=\!-0.7797$. For the 100\,GeV/nucleon beam, the hadronic spin-flip amplitude correction to such measured beam polarization is about $-0.8\%$ and the correction due to the $m^2/s$ term in (\ref{eq:kappa'}) is about $+0.8\%$.

The hadronic spin-flip dependent uncertainties in the beam polarization measurement may be evaluated as
\begin{align}
  \delta\xi_0 &= \pm0.1\,I_5^{pp} + G_\varkappa\,\delta{I_5^{pp}}
  \approx \pm0.5\%,   
  \label{eq:dPmeth} \\
   G_\varkappa &= 2\times\left[0.27/\varkappa_h'-1/\varkappa_p'\right]=-1.50.
  \label{eq:G}
\end{align}
The numerical estimate of $\delta\xi_0$ follows from the quadratic sum of the uncertainties in the ratios $r_5^{hp,ph}/r_5^{pp}$ and in the value of $I_5^{pp}$. For the calculation, Run\,15 value of $I_5^{pp}\!=\!-0.0053$ but Run\,17 evaluation of $\delta{I_5^{pp}}\!=\!\pm0.0035$ were used. Since some essential sources of systematic errors were eliminated after Run\,15 (100\,GeV), the 255\,GeV value of  $\delta{I_5^{pp}}$ is more relevant in this paper. However, it is assumed that better measurements of $I_5$ at 100\,GeV will be done in a future RHIC Run\,24 or at EIC.

Pedantically,  $I_5$ and $R_5$ in Eq.\,(\ref{eq:He3Pol}) should be replaced by\,\cite{Buttimore:1998rj}
\begin{equation}
  I_5 \to I_5-\delta_CR_5\quad\text{and}\quad R_5\to R_5-\rho I_5.
\end{equation}
However, such corrections are inessential in the context of the EIC requirement (\ref{eq:systEIC}).

A small correction to Eq.\,(\ref{eq:em5}) due to the difference between Dirac and Pauli form factors\,\cite{Poblaguev:2019vho} may be included in $\omega_\varkappa(T_R)$ [Eq.\,(\ref{eq:kappaCorr})] and, thus, does not require special consideration. 

  Estimating $r_5^{ph}$ (\ref{eq:r5ph}) and $r_5^{hp}$ (\ref{eq:r5hp}), it was assumed that hadronic spin-flip parameters $r_5^{pn}\!=r_5^{np}\!=\!r_5^{pp}$ are the same for proton-proton and proton-neutron scattering. This equity can be violated by the isovector component of the scattering amplitude. For the C-odd, J=1 Regge pole exchange, the coupling ratio for ${np}$ and ${pp}$ for nonflip amplitudes is\,\cite{PDG:2017pp}
  \begin{equation}
    f^-_{np}/f^-_{pp} = 0.901\pm0.023.
  \end{equation}
  Assuming the same ratio for the spin-flip amplitudes and using the experimentally evaluated Reggeon spin-flip couplings\,\cite{Poblaguev:2019saw}, one finds
  \begin{equation}
    \delta\xi_0^{pn} \approx-0.22\%,\qquad    
    \delta\xi_1^{pn} \approx-0.19\%.
  \end{equation}
The correction to $\delta\xi_0$ is small compared to the uncertainty given in Eq.\,(\ref{eq:dPmeth}) and, thus, can be discarded.
  
   At HJET, systematic uncertainty in the recoil proton energy calibration was evaluated\,\cite{Poblaguev:2020qbw} to be
  \begin{align}
    \delta^\text{cal}T_R = &\delta{c_0} + \delta{c_1}\,T_R,  \label{eq:dc1}\\
    &\delta{c_0} = \pm15\,\text{keV},~~\delta{c_1} = \pm0.01
  \end{align}
  The constant term can contribute to the $\sigma_P^\text{syst}/P$, but the resulting value,
  \begin{align}
    \delta\xi_0^{T_R} =
    &G_\varkappa R_5^{pp} \delta{c_0}/T_c^{hp} \approx \pm0.06\%, 
  \end{align}
   can be neglected.

  \section{Summary}

  I have studied feasibility of a polarized hydrogen gas jet target to be used for precision determination of the ${}^3\text{He}^\uparrow$ beam polarization at EIC. The numerical estimates were done for 100\,GeV/nucleon beam energy.

  It was shown that concurrent measurement of the ${}^3\text{He}$ beam $a_\text{N}^{h}(T_R)$ and hydrogen jet target $a_\text{N}^{p}(T_R)$ spin correlated asymmetries dependence on the recoil proton kinetic energy $T_R$ allows one to precisely determine the beam polarization. For that, the measured, almost linear, function of $T_R$, defined as
\begin{align}
  &P_\text{meas}(T_R) = P_\text{jet} \frac{a_\text{N}^{h}(T_R)}{a_\text{N}^{p}(T_R)}%
  \nonumber \\
  &\quad\;\;
  \times\frac{\varkappa_p-m_p^2/m_hE_\text{beam}-2I_5-2R_5\,T_R/T_c}%
       {\varkappa_h-m_h/E_\text{beam}    -0.54I_5-0.54R_5\,T_R/T_c},
       \label{eq:He3_Pol}
\end{align}
should be extrapolated to $T_R\!\to\!0$:
\begin{equation}
 \!\!\!P_\text{beam}\!=\!P_\text{meas}(0)\times%
  \left[1\pm0.006_\text{syst}\pm0.005_{r_5}\pm0.002_\text{mod}\right]\!.
  \label{eq:PhErrs}
\end{equation}
 Here, $\varkappa_p\!=\!1.793$, $\varkappa_h\!=\!-1.398$,
  $E_\text{beam}$ is the helion beam energy per nucleon, $T_c\!\approx\!0.7\,\text{MeV}$ (for 100\,GeV/nucleon helion beam), and $r_5\!=\!R_5\!+\!iI_5$ is the elastic $\mathit{pp}$ hadronic spin-flip amplitude parameter determined at $s\!=\!2m_pE_\text{beam}$.

For the measurement-dependent systematic error in Eq.\,(\ref{eq:PhErrs}), I used results of emulation of HJET performance at EIC\,\cite{Poblaguev:2020Og}, the value of $r_5^{pp}$ related uncertainty is based on Run\,17 (255\,GeV) proton beam measurements at RHIC\,\cite{Poblaguev:2019saw}, and the model-dependent uncertainties in the $r_5^{pp}$ extrapolation to $r_5^{ph},\,r_5^{hp}$ were estimated in this paper.

The evaluated accuracy of the beam polarization measurement meets the EIC requirement $\sigma_P^\text{syst}/P\!\lesssim\!1\%$ for hadronic polarimetry. Remarkably that breakup correction $b^{hp}(T_R)$, which, potentially, may give an inappropriately (for the polarimetry) large contribution to $A_\text{N}$, cancels in Eq.\,(\ref{eq:He3_Pol}). 

Possible errors in predetermined values of the total cross section $\delta\sigma_\text{tot}^{hp}$, energy calibration $\delta{c_1}$\,(\ref{eq:dc1}), and absorptive correction $\delta{\cal B}_a^\text{sf}$\,(\ref{eq:r5Meas}) may result in $P_\text{meas}(T_R)$ dependence on $T_R$ 
\begin{align}
  T_c\,dP_\text{meas}&(T_R)/{dT_R } = \nonumber\\
  &G_\varkappa\!\times\!\left[%
  {\delta \sigma_\text{tot}^{hp}}/{\sigma_\text{tot}^{hp}} +
  \delta{c_1} +
  \delta{\cal B}_a^\text{sf}\varkappa_p/2
  \right]\!.
  \label{eq:T/Tc}
\end{align}
So, the sum of errors in values of $\sigma_\text{tot}^{hp}$, $c_1$, and ${\cal B}_a^\text{sf}$  defined in Eq.\,(\ref{eq:T/Tc}) can be experimentally evaluated  with an accuracy of ${\cal O}(10^{-3})$  directly in the linear fit of $P_\text{beam}^h(T_R)$.

The HJET (with possible improvements if required) is being considered for the measurement of the proton beam absolute polarization at EIC. According to estimates done in this paper, exactly that
polarimeter also has the ability to precisely measure helion beam polarization (if proton-proton $r_5$ for the beam energy used is already known). Therefore, some common potential problems for the EIC proton and helion beam polarimetry, e.g., due to much shorter bunch spacing of 10\,ns at EIC compared to 107\,ns at RHIC\,\cite{Poblaguev:2020Og}, were not considered here.

The ${}^3\text{He}$ beam polarization uncertainty dependence on $r_5^{pp}$ can be reduced by a factor of about 4 if an {\em unpolarized} hydrogen jet target will be used. In this case, also, event statistics can be significantly increased snd some essential systematic uncertainties in the measurements can be eliminated. However, to implement this method much better knowledge of the $d\sigma^{hp}/dt$ parametrization (including the beam ${}^3\text{He}$ breakup effect) is needed.

  It should be also noted that developing alternative approaches to ${}^3\text{He}$ beam polarimetry at EIC, e.g., polarized ${}^3\text{He}$ target, is important to obtain a confident result.

  \acknowledgments{
    The author would like to thank N.~H.~Buttimore, B.~Z.~Kopeliovich, and A.~Zelenski for useful discussions and valuable comments, and acknowledges support from the Office of Nuclear Physics in the Office of Science of the US Department of Energy. This work is authored by employees of Brookhaven Science Associates, LLC under Contract No. DE-SC0012704 with the U.S. Department of Energy.
}

\end{document}